%% file: PhysicsObjects-TopCMS.tex
\documentclass[12pt]{article}
\usepackage{graphicx}

\newcommand\pubnumber{SNSN-323-63}
\newcommand\pubdate{\today}

\def\institute{on behalf of the CMS collaboration\\
Affiliation:\\
Centre for High Energy Physics, \\
Indian Institute of Science, Bangalore - 560012, India\\
E-mail: jyothsna@cern.ch}
\def\speaker{\footnote{Speaker }}

\def\Title#1{\begin{center} {\Large #1 } \end{center}}
\def\Author#1{\begin{center}{ \sc #1} \end{center}}
\def\Address#1{\begin{center}{ \it #1} \end{center}}

\newcommand\pubblock{\rightline{\begin{tabular}{l} \pubnumber\\
         \pubdate  \end{tabular}}}
\newenvironment{Abstract}{\begin{quotation}  }{\end{quotation}}
\newenvironment{Presented}{\begin{quotation} \begin{center} 
             PRESENTED AT\end{center}\bigskip 
      \begin{center}\begin{large}}{\end{large}\end{center} \end{quotation}}

\input econfmacros.tex

\begin{document}
\begin{titlepage}
\pubblock

\vfill
\Title{Physics objects for top physics in CMS}
\vfill
\Author{Jyothsna Rani Komaragiri\speaker}
\Address{\institute}
\vfill
\begin{Abstract}
We present the status and performance of the physics objects with data corresponding to an integrated luminosity of 12.9 fb$^{-1}$ collected at 13 TeV by 
the CMS experiment. We cover the performance of the physics objects extensively used in physics analyses involving top quarks.
\end{Abstract}
\vfill
\begin{Presented}
$9^{th}$ International Workshop on Top Quark Physics,\\
Olomouc, Czech Republic,  September 19--23, 2016
\end{Presented}
\vfill
\end{titlepage}
\def\thefootnote{\fnsymbol{footnote}}
\setcounter{footnote}{0}
%


\section{Introduction}

The Large Hadron Collider (LHC) is providing proton-proton collisions at an unprecedented center-of-mass energy of $\sqrt{s}$ = 13 TeV with the start of Run 2 
in 2015. The collision data recorded by the Compact Muon Solenoid (CMS)~\cite{cms} experiment reopen the opportunity to search for new 
physics phenomena at the TeV scale and to perform precision measurements of the standard model at a higher center-of-mass energy. To carry out these searches 
and precision measurements we need to understand our detector and the performance of various physics objects at higher center-of-mass energy as well as varying 
running conditions (high pile up, increasing instantaneous luminosity). The decay products from a top quark contain most of the physics objects we identify in 
our detector. We present the performances of all physics objects used in the physics analyses involving top quarks.

\section{Event Reconstruction}

Stable particles are identified with the Particle Flow (PF) algorithm~\cite{pf} that reconstructs each individual particle with an optimized combination of 
information from the various elements of the CMS detector. The energy of photons is directly obtained from the ECAL measurement, corrected for zero-suppression 
effects. The energy of electrons is determined from a combination of the electron momentum at the primary interaction vertex as determined by the tracker, the 
energy of the corresponding ECAL cluster, and the energy sum of all bremsstrahlung photons spatially compatible with originating from the electron track. The 
energy of muons is obtained from the curvature of the corresponding track. The energy of charged hadrons is determined from a combination of their momentum 
measured in the tracker and the matching ECAL and HCAL energy deposits, corrected for zero-suppression effects and for the response function of the calorimeters 
to hadronic showers. Finally, the energy of neutral hadrons is obtained from the corresponding corrected ECAL and HCAL energy.

\section{Jets and Missing Transverse Energy}

\begin{figure}[htb] 
\centering
\includegraphics[height=2.2in, width=2.2in]{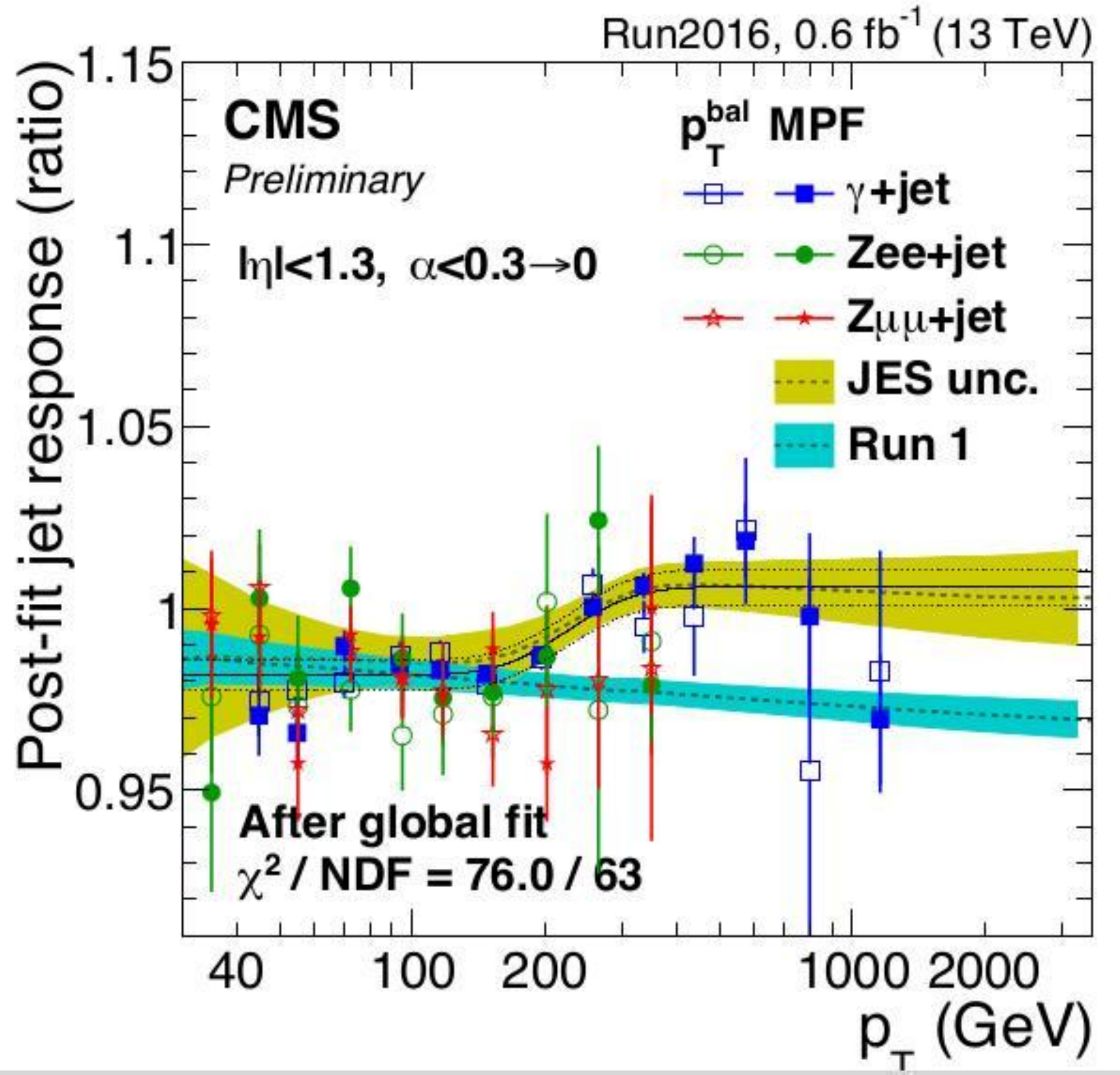}
\includegraphics[height=2.2in, width=2.2in]{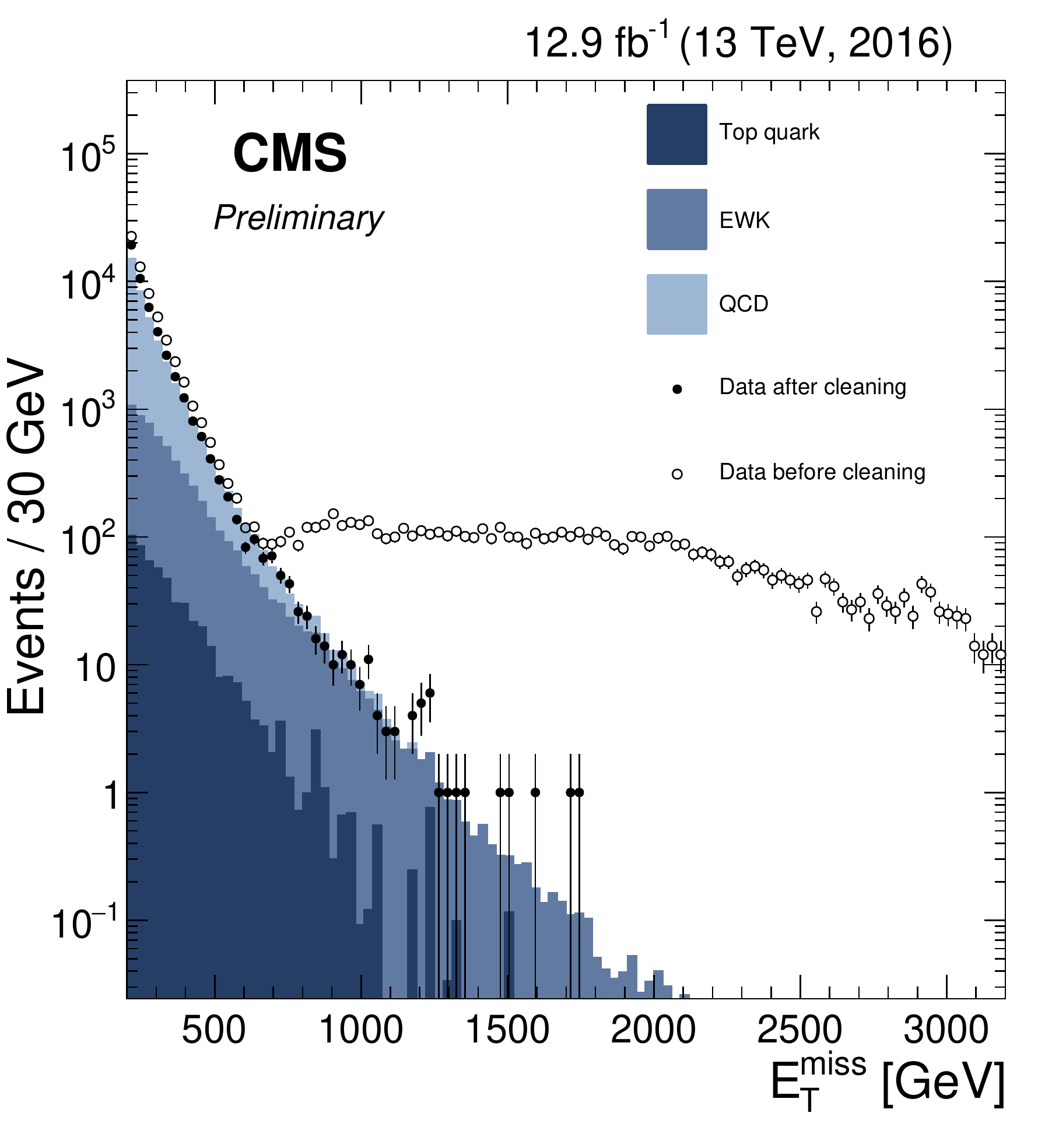}
\caption{Data and simulation comparison	for the jet response dependency	on the jet $p_T$ (left).
The MissingET distributions for events passing the dijet selection with the 2016 cleaning algorithms applied including the one based on jet identification 
requirements (filled markers), without the 2016 cleaning algorithms applied (open markers), and from simulation (filled histograms) (right).}
\label{fig:jetmet}
\end{figure}

Jets are reconstructed by clustering PF particles using the anti-kT (AK) jet clustering algorithm, with a distance parameter R = 0.4 (AK4 jets). For the boosted 
topologies, jets are clustered with a larger opening angle corresponding to R = 0.8 (AK8 jets). When clustering the particles in jets, isolated electrons and 
muons as well as charged particles associated with other interaction vertices are removed. Jet momentum is determined as the vectorial sum of all particle 
momenta in the jet. Jet energies are calibrated to correct for the different detector response as a function of the transverse momentum and pseudorapidity of 
the jets. Furthermore, an offset correction is applied to jet energies to take into account the contribution from additional proton-proton interactions within 
the same bunch crossing. Comparison of jet response dependency on the jet $p_T$ in data and simulation  is shown Figure~\ref{fig:jetmet} (left). Jet energy
corrections and resolution can be found in Ref.~\cite{jet}.

The default missing transverse energy in the transverse plane (MET) is reconstructed by the vector sum of all particle-flow particles in the opposite direction.
Performance of 2016 cleaning algorithms on missing transverse energy distribution is shown in Figure~\ref{fig:jetmet} (right). More details and performance of MET
algorithms can be found in Ref.~\cite{met}.

\section{b tagging}

\begin{figure}[htb]
\centering
\includegraphics[height=2.2in, width=2.2in]{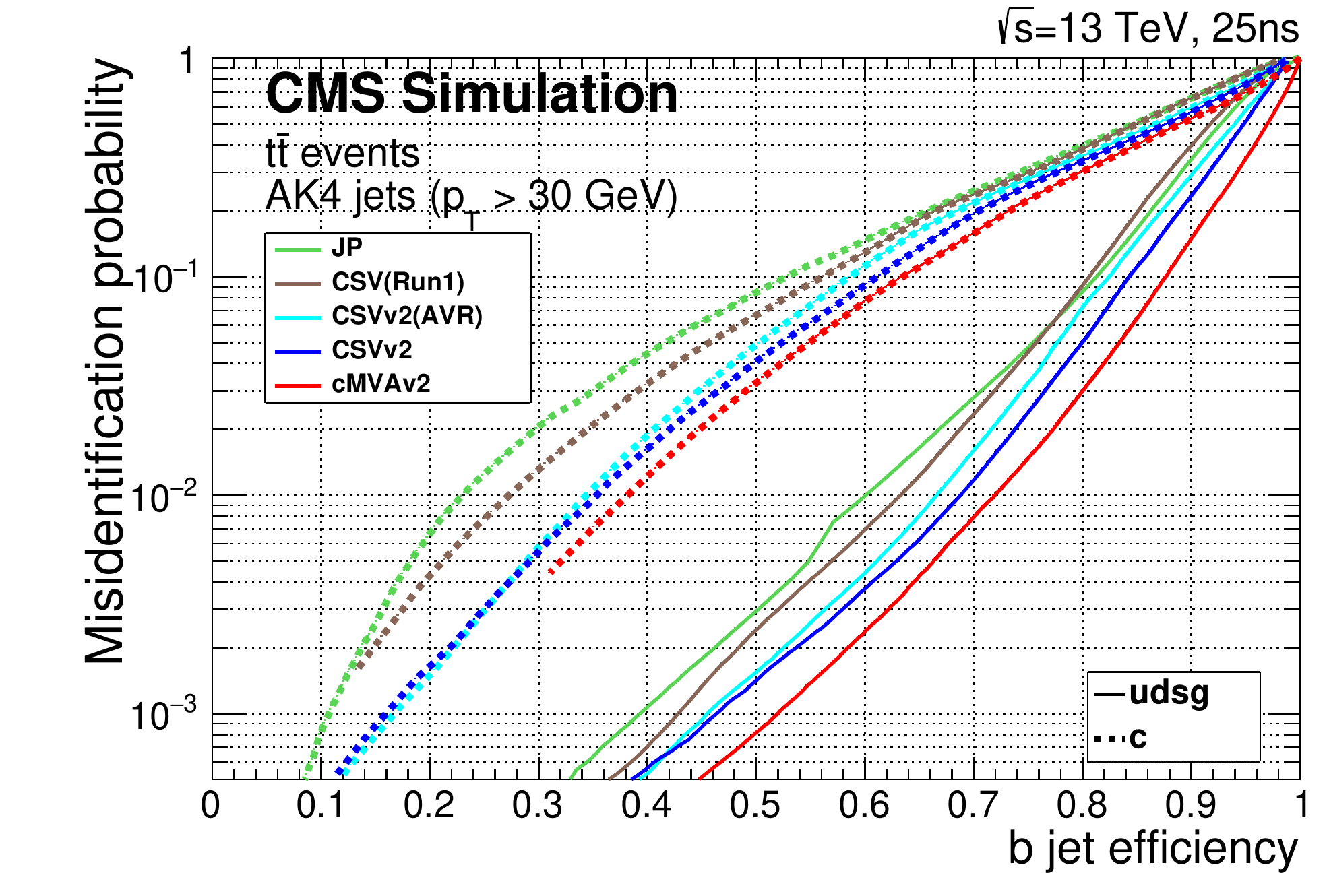}
\includegraphics[height=2.2in, width=2.2in]{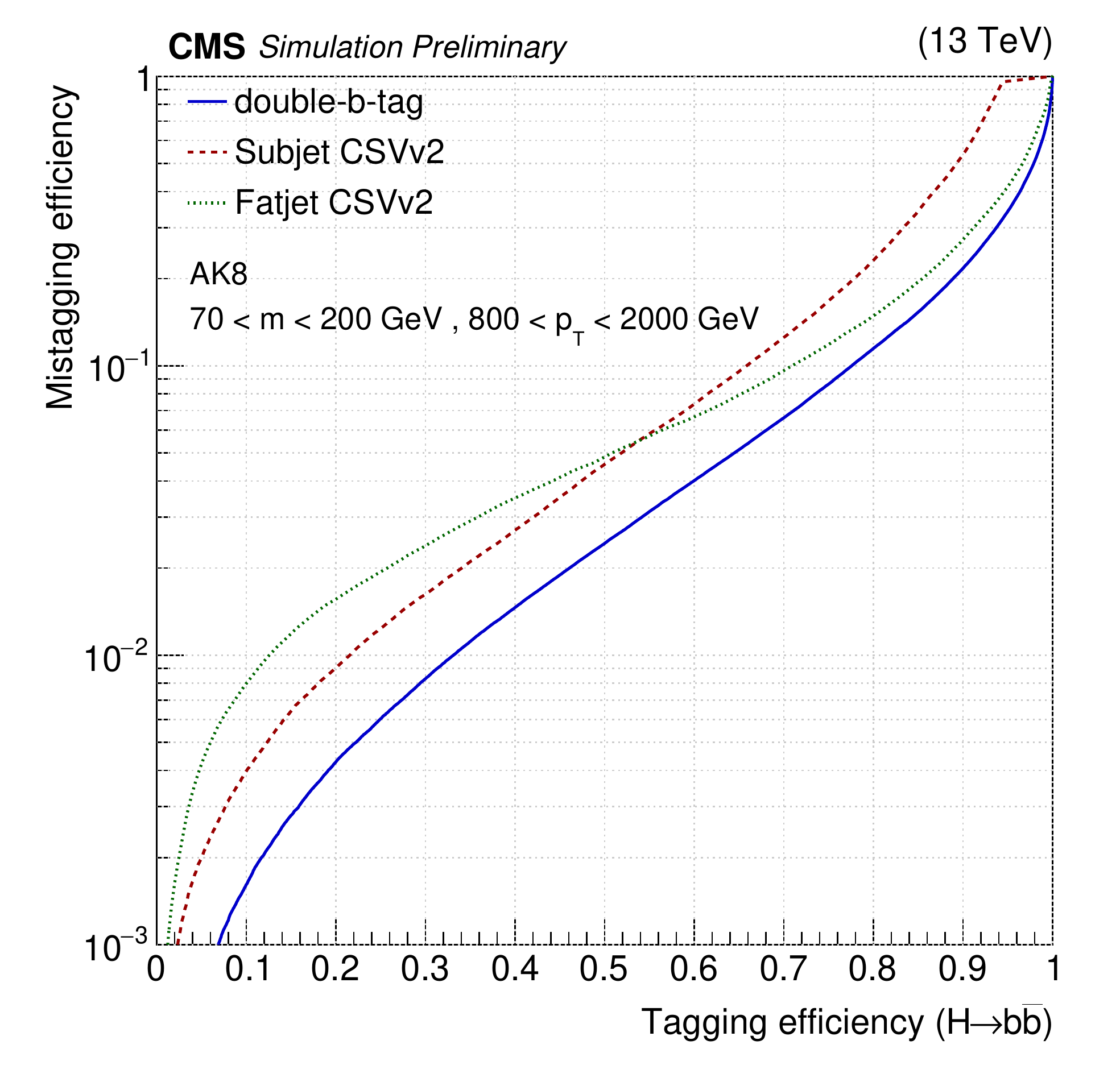}
\caption{Performance of the b jet identification efficiency algorithms demonstrating the probability for non-b jets to be misidentified as b jet as a 
function of the efficiency to correctly identify b jets (left). Comparison of the performance of the double-b tagger, the minimum CSVv2 value among the 
two subjets b tag scores, and fat jet b tag which exploits CSVv2 algorithm (right). }
\label{fig:btag}
\end{figure}

Details about b tagging in CMS can be found in~\cite{btag}. The Combined Secondary Vertex (CSVv2) tagger using multivariate technique based on track 
information and vertex information is the best performed algorithm in CMS. The other available taggers are Jet Probability (JP) tagger based on the track information 
and combined Multivariate Analysis (cMVAv2) tagger which makes use of track, vertex and soft lepton information using multivariate technique. The performance of 
the JP, CSVv2 and cMVAv2 taggers in simulated top pair production events is shown in Figure~\ref{fig:btag} (left). The performance of the b tagging algorithms has 
been measured using the data. The data and simulation comparisons can be found in Ref.~\cite{btag1}.

We also have a c tagging algorithm for identifying jets originating from c-quarks. More details about c tagger along with its performance can be found 
in Ref.~\cite{ctag}.


In boosted topologies b tagging can be performed applying the standard CSVv2 tagger to larger cone jets (know as fatjets) or to their subjets, obtained with 
substructure algorithms. A dedicated training has also been performed for fat jets originating from a boosted resonance decaying into a $b\bar{b}$ pair, 
outperforming the subjet and fatjet b tagging in such a topology, as shown in Figure~\ref{fig:btag} (right).

\section{Electrons}
\begin{figure}[htb]
\centering
\includegraphics[height=2.2in, width=2.2in]{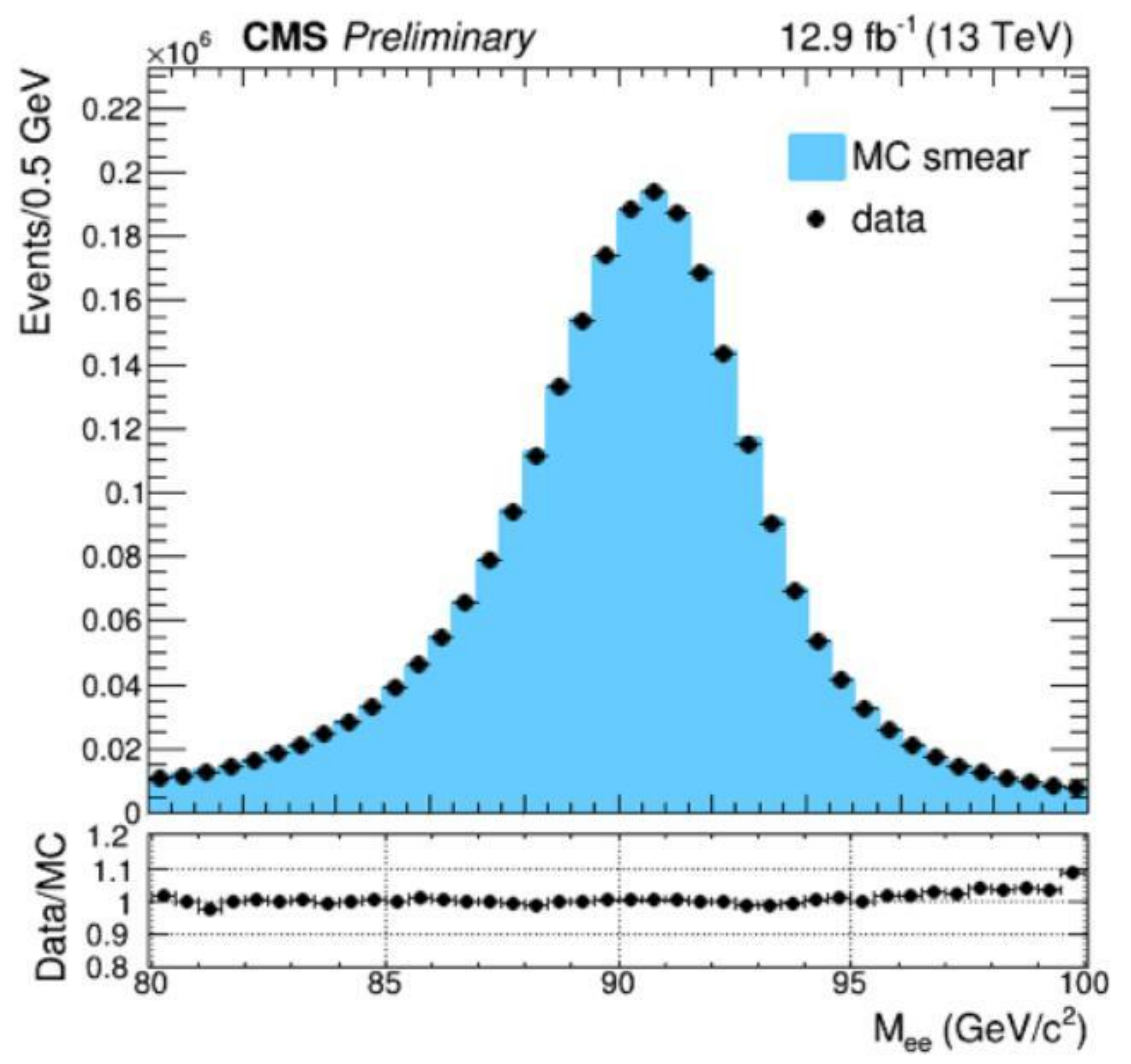}
\includegraphics[height=2.2in, width=2.2in]{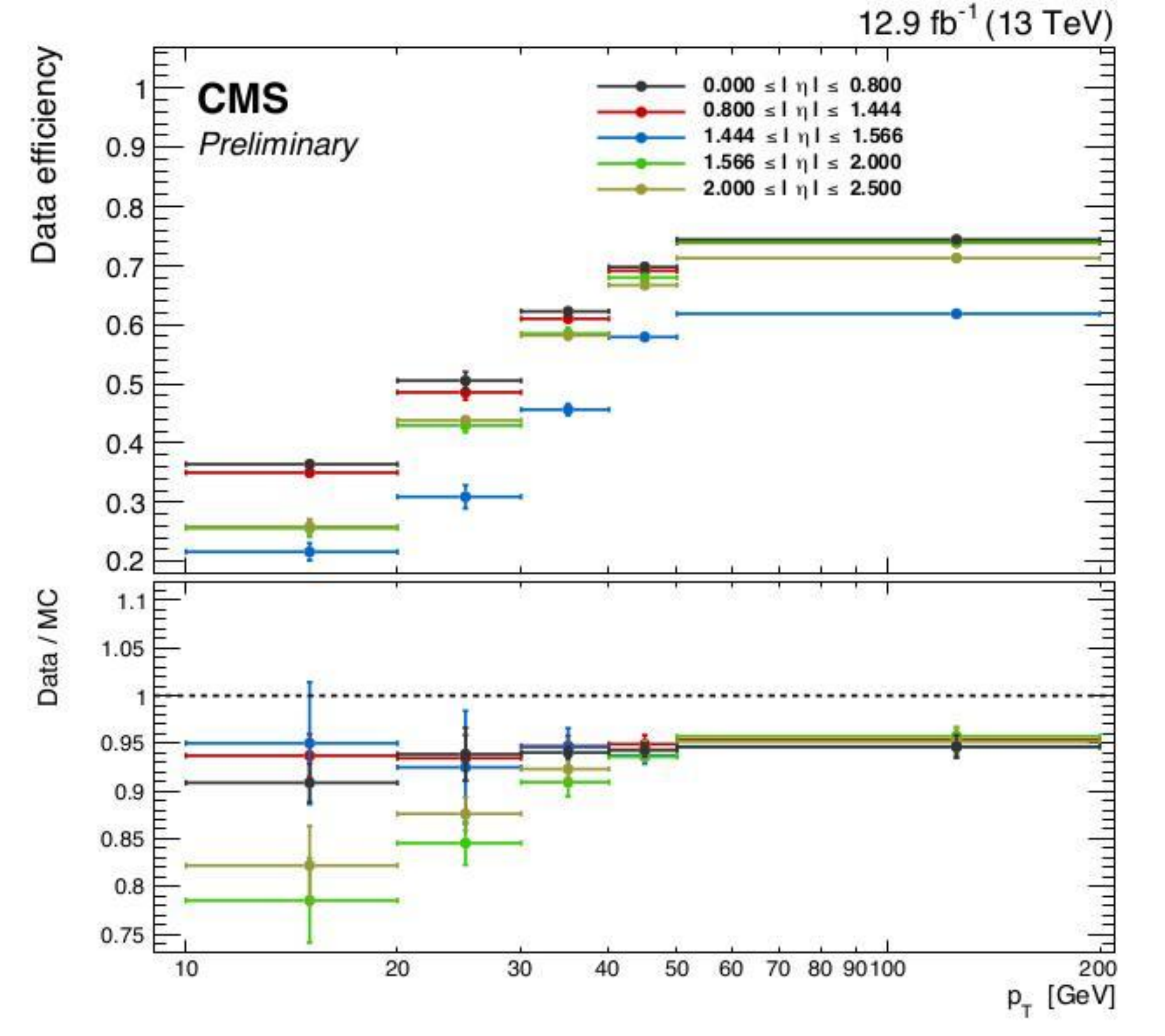}
\caption{Data (black dots) and simulation (light blue) $Z \rightarrow ee$ mass distribution reconstructed with electron tuned energy corrections for barrel-barrel 
electron pairs (left). Electron identification efficiency in data (right top) and data to simulation efficiency ratios (right bottom) measured for the tight 
cut-based identification.}
\label{fig:elefig}
\end{figure}

Electron reconstruction in Run2 is completely integrated with particle flow (PF) and provides additional flexibility at analysis level. As can be seen in 
Figure~\ref{fig:elefig} (left) reconstructed $Z \rightarrow ee$ mass distribution using electron tuned energy corrections for barrel-barrel electron pairs shows 
good agreement between data and simulation. The efficiency for cut-based electron identification is measured with the tag and probe method and shown in five 
pseudorapidity ranges as a function of the electron transverse momentum and the results are shown Figure~\ref{fig:elefig} (right). Other kinematics 
distributions of electrons are also compared between data and simulation and show a good agreement (see Ref.~\cite{ele}).

\section{Muons}
\begin{figure}[htb]
\centering
\includegraphics[height=2.2in, width=2.2in]{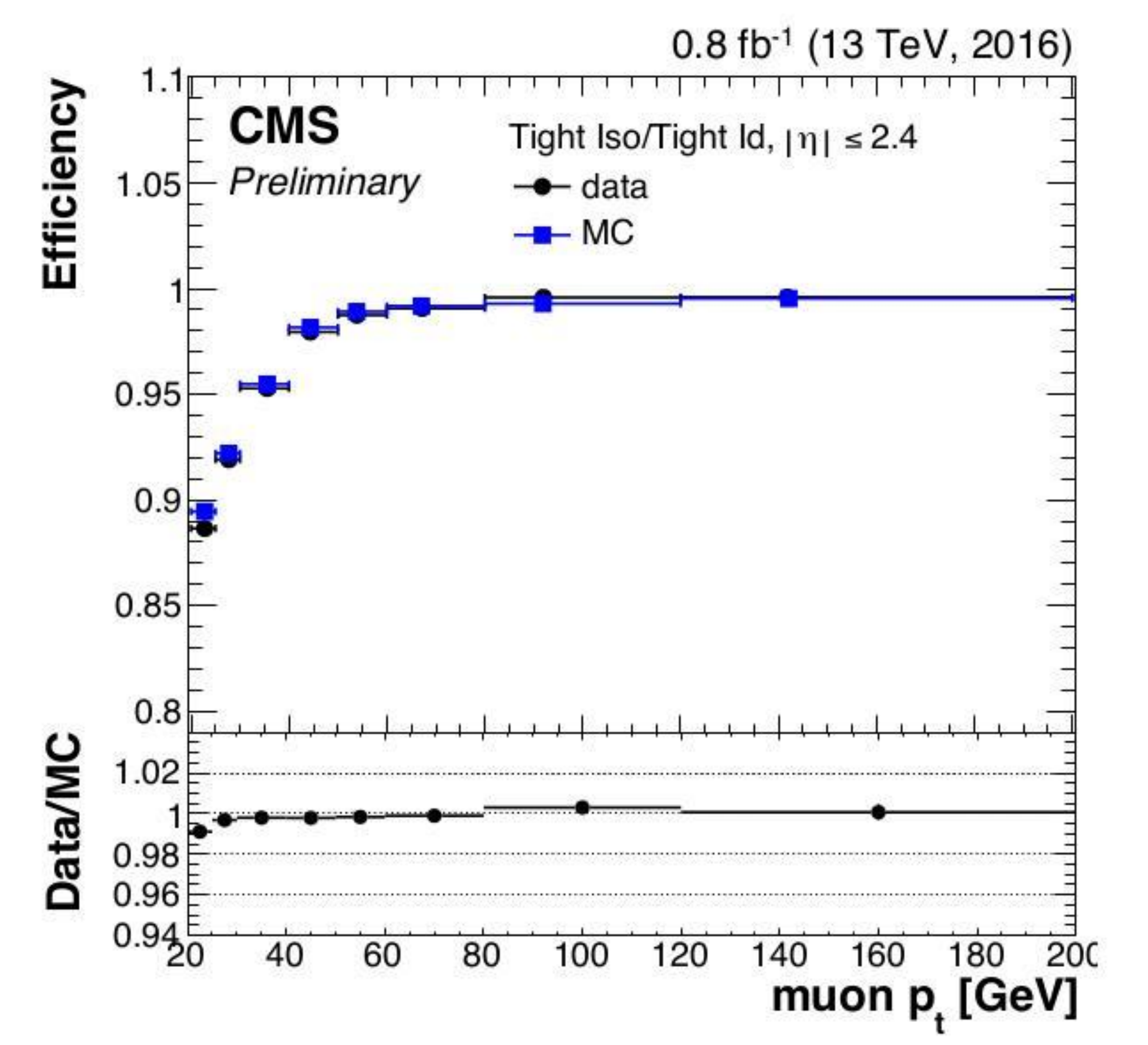}
\includegraphics[height=2.2in, width=2.2in]{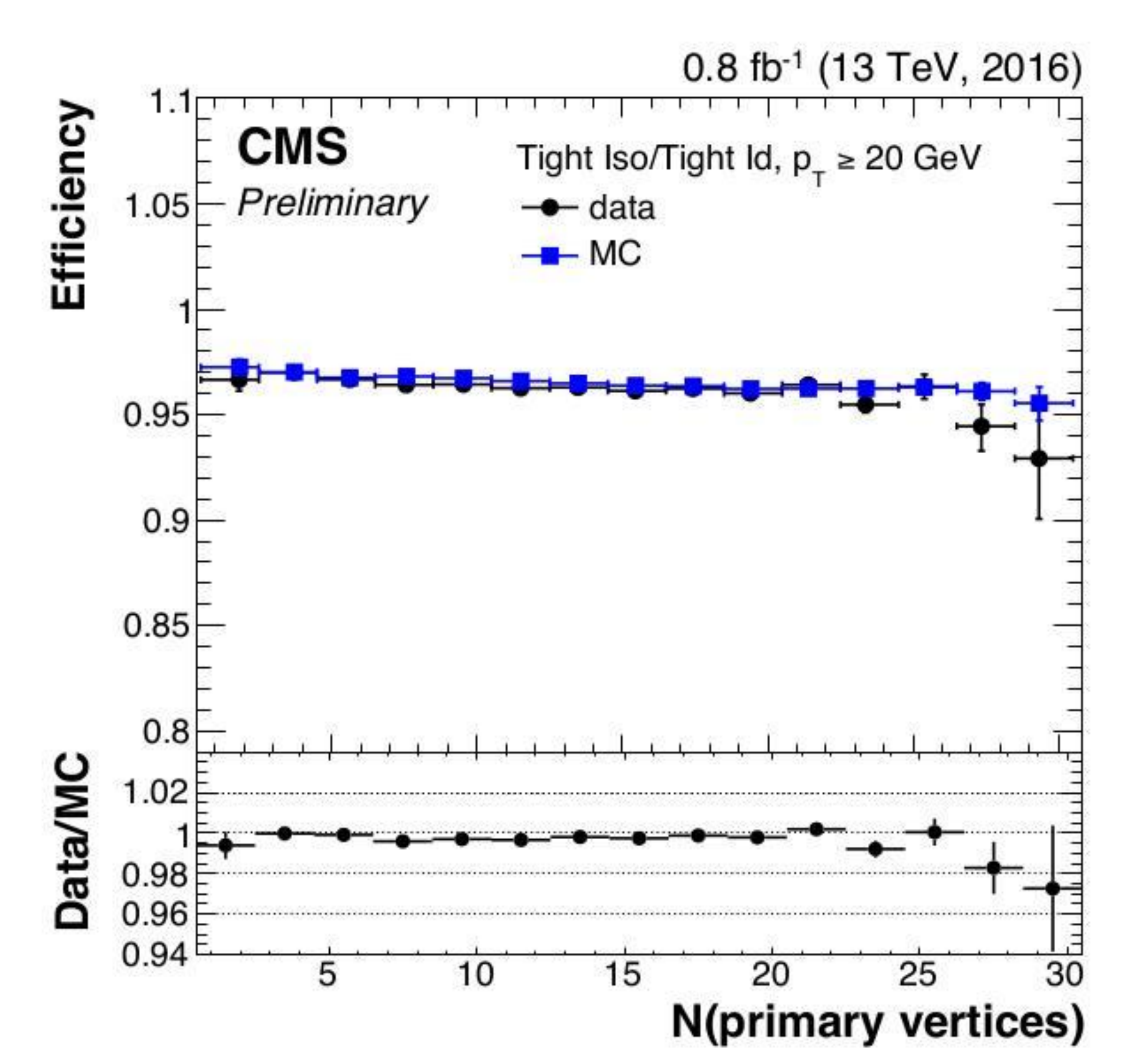}
\caption{Muon identification and isolation efficiency for tight working points in shown as function of probe muon $p_T$ (left) and number of primary vertexes (right).}
\label{fig:mufig}
\end{figure}

Efficiencies are computed by means of the tag-and-probe method exploiting $Z \rightarrow \mu \mu$ resonances. Computed identification and isolation efficiencies for
tight working point as a function of probe muon $p_T$ (left) and number of primary vertices (right) are shown in Figure~\ref{fig:mufig}. Good agreement between data 
and simulation can be seen. Other kinematics distributions of muons are also compared between data and simulation and show a good agreement (see Ref.~\cite{mu}).


\section{Conclusions}
The object performance study is the key for the measurements involving top quarks and also searches beyond the Standard Model. Performance of various 
physics objects extensively used in Top physics at CMS using RunII data shows good agreement between data and simulation.



\end{document}

%% file: econfmacros.tex
\def\beq{\begin{equation}}
\def\eeq#1{\label{#1}\end{equation}}
\def\eeqn{\end{equation}}

\def\beqa{\begin{eqnarray}}
\def\eeqa#1{\label{#1}\end{eqnarray}}
\def\eeqan{\end{eqnarray}}

\let\bar=\overbar

\def\Dslash{\not{\hbox{\kern-4pt $D$}}}
\def\dslash{\not{\hbox{\kern-2pt $\del$}}}

\def\msb{{\bar{\ssstyle M \kern -1pt S}}}

